\documentclass[prd,twocolumn,a4paper,superscriptaddress,floatfix]{revtex4}
\usepackage{graphicx}
\usepackage{bbm}
\usepackage{amssymb}
\usepackage{calligra}
\usepackage{lipsum}
\usepackage[T1]{fontenc}
\DeclareFontShape{T1}{calligra}{m}{n}{<->s*[2.5]callig15}{}
\DeclareMathAlphabet{\mathcalligra}{T1}{calligra}{m}{n}
\DeclareMathAlphabet{\mathpzc}{OT1}{pzc}{m}{it}

\begin{document}
%My commands
\newcommand{\be}{\begin{equation}}
\newcommand{\ee}{\end{equation}}
\newcommand{\bq}{\begin{eqnarray}}
\newcommand{\eq}{\end{eqnarray}}
\newcommand{\bsq}{\begin{subequations}}
\newcommand{\esq}{\end{subequations}}
\newcommand{\bc}{\begin{center}}
\newcommand{\ec}{\end{center}}
\newcommand\lapp{\mathrel{\rlap{\lower4pt\hbox{\hskip1pt$\sim$}} \raise1pt\hbox{$<$}}}
\newcommand\gapp{\mathrel{\rlap{\lower4pt\hbox{\hskip1pt$\sim$}} \raise1pt\hbox{$>$}}}
\newcommand{\dpar}[2]{\frac{\partial #1}{\partial #2}}
\newcommand{\sdp}[2]{\frac{\partial ^2 #1}{\partial #2 ^2}}
\newcommand{\dtot}[2]{\frac{d #1}{d #2}}
\newcommand{\sdt}[2]{\frac{d ^2 #1}{d #2 ^2}}
\newcommand{\vv}[0]{{\bar v}}
\newcommand{\ave}[1]{\left< #1 \right>}

\title{The Stochastic Gravitational Wave Background Generated by Cosmic String Networks: the Small-Loop Regime}

\author{L. Sousa}
\email[Electronic address: ]{Lara.Sousa@astro.up.pt}
\affiliation{Centro de Astrof\'{\i}sica da Universidade do Porto, Rua das Estrelas, 4150-762 Porto, Portugal}
\affiliation{Departamento de F\'{\i}sica e Astronomia da Faculdade de Ci\^encias
da Universidade do Porto, Rua do Campo Alegre 687, 4169-007 Porto, Portugal}

\author{P.P. Avelino}
\email[Electronic address: ]{ppavelin@fc.up.pt}
\affiliation{Centro de Astrof\'{\i}sica da Universidade do Porto, Rua das Estrelas, 4150-762 Porto, Portugal}
\affiliation{Departamento de F\'{\i}sica e Astronomia da Faculdade de Ci\^encias
da Universidade do Porto, Rua do Campo Alegre 687, 4169-007 Porto, Portugal}
\affiliation{Sydney Institute for Astronomy, School of Physics, A28, The University of Sydney, NSW 2006, Australia}

\begin{abstract}
We consider an alternative approach for the computation of the stochastic gravitational wave background generated by small loops produced throughout the cosmological evolution of cosmic string networks and use it to derive an analytical approximation to the corresponding power spectrum. We show that this approximation produces an excellent fit to more elaborate results obtained using the Velocity-dependent One-Scale model to describe cosmic string network dynamics, over a wide frequency range, in the small-loop regime.
\end{abstract} 
\maketitle

\section{Introduction\label{intro}}
Cosmic strings networks may be produced as a consequence of symmetry-breaking phase transitions \cite{Kibble:1976sj}, being a crucial prediction of many grand-unified scenarios. These networks may survive throughout the cosmological history, potentially leaving behind a variety of observational signatures (see e.g. \cite{Vilenkin:1984ea,Avelino:1997hy,Avelino:1998vu,Sarangi:2002yt,Avelino:2003nn,Bevis:2010gj} and references therein). One such signature is the stochastic gravitational wave background (SGWB) generated by string loops created as a result of string interactions. These loops radiate their energy in gravitational waves (GWs) and their emissions generate a characteristic SGWB \cite{Vilenkin:1981bx,Hogan:1984is,Brandenberger:1986xn,Accetta:1988bg}.

The SGWB power spectrum generated by cosmic string networks may be probed using diverse astrophysical experiments: GW detectors \cite{Sigg:2008zz,Accadia:2011zzc,Kuroda:2010zzb,AmaroSeoane:2012km,Kawamura:2011zz}), pulsar timing experiments \cite{Manchester:2007mx,Ferdman:2010xq,Demorest:2012bv}, small-scale fluctuations and B-mode polarization of CMB \cite{Smith:2006nka,Planck:2006aa,Baumann:2008aq}); and big-bang nucleosynthesis \cite{Cyburt:2004yc}. There is thus the prospect either for the detection of specific cosmic string signatures in the SGWB or for the tightening of current constraints on string tension. It is, therefore, important to accurately characterize the SGWB spectrum and to understand its dependence on the large-scale properties of string networks and on the size and emission spectrum of the loops. There are, in the literature, several computations of the SGWB spectrum \cite{Caldwell:1991jj,Caldwell:1996en,Battye:1997ji,Damour:2001bk,Damour:2004kw,Siemens:2006vk,Hogan:2006we,Polchinski:2006ee,DePies:2007bm,Olmez:2010bi,Binetruy:2012ze,Kuroyanagi:2012wm,Sanidas:2012ee,Sousa:2013aaa} based on different assumptions about string network dynamics. In this paper, we present an alternative method to compute the SGWB generated by a realistic cosmic string network and we use it to derive an analytical approximation to the SGWB power spectrum, over a wide frequency range, in the small-loop regime.

\section{Cosmic string network evolution\label{string}}
The Velocity-dependent One-Scale (VOS) model \cite{Martins:1996jp,Martins:2000cs} describes the time evolution of the characteristic lengthscale of the network, $L$, and of its root-mean-square (RMS) velocity, $\vv$, thus allowing for a quantitative characterization of string network dynamics. If one assumes that the cosmic string network is statistically homogeneous on sufficiently large scales, one may define its characteristic lengthscale as $L\equiv(\mu/\rho)^{1/2}$, where $\mu$ is the cosmic string tension, and $\rho$ is the average energy density of long strings. In the limit of infinitely thin cosmic strings, the following evolution equations for $L$ and $\vv$ can be obtained by averaging the microscopic Nambu-Goto equations of motion \cite{Martins:1996jp,Martins:2000cs} (see also \cite{Sousa:2011ew,Sousa:2011iu} for a more general derivation of the VOS equations):
\bq
2\frac{dL}{dt} & = & \left(2H+\frac{\vv^2}{\ell_d}\right)L\,,\label{vosL}\\
\frac{d\vv}{dt} & = & (1-\vv^2)\left[\frac{k(\vv)}{L}-\frac{\vv}{\ell_d}\right]\,,\label{vosv}
\eq
where $H=\dot{a}/a$ is the Hubble parameter, $a$ is the cosmological scale factor and dots represent derivatives with respect to physical time. We have also introduced the damping lengthscale, $\ell_d^{-1}=2H+\ell_f^{-1}$, that accounts for the damping caused by the Hubble expansion and also for the effect of frictional forces caused by interactions with other fields (encoded in the frictional lengthscale, $\ell_f$). We shall assume for the remainder of this paper that $\ell_f=\infty$. Moreover,
\be
k(\vv)=\frac{2\sqrt{2}}{\pi}\left(1-\vv^2\right)\left(1+2\sqrt{2}\vv^3\right)\frac{1-8\vv^6}{1+8\vv^6}
\label{kpar}
\ee
is an adimensional curvature parameter that encodes the effects caused by the small-scale structure on long strings (see \cite{Martins:2000cs}).

Cosmic string intersections often result in the formation of loops that detach from the long string network. The energy lost into loops by this network can be written as \cite{Kibble:1984hp}
\be
\left.\frac{d\rho}{dt}\right|_{\rm loops}={\tilde c}{\bar v}\frac{\rho}{L}\,,
\label{loss}
\ee
where ${\tilde c}$ is a phenomenological parameter that characterizes the efficiency of the loop-chopping mechanism. Numerical simulations indicate that ${\tilde c}=0.23\pm 0.04$ is a good fit both in the matter and radiation eras \cite{Martins:2000cs}.

These loops start decaying radiatively once they detach from the cosmic string network and, thus, they have a finite lifespan. Consequently, the network loses energy at the rate given by Eq. (\ref{loss}). This effect is included in the VOS equations by adding the following term to the right-hand side of Eq. (\ref{vosL})
\be
\left.\frac{dL}{dt}\right|_{loops}=\frac{1}{2}\tilde{c}\vv\,.
\label{lossL}
\ee

Eqs. (\ref{vosL}), (\ref{vosv}) and (\ref{lossL}) are the basis of the VOS model and they describe the large-scale evolution of cosmic string networks. Interestingly, the linear scaling regime \cite{Bennett:1987vf,Albrecht:1989mk,Allen:1990tv,Copeland:1991kz,Vincent:1996rb} arises naturally in this model. Indeed, a regime of the form 
\be
\frac{L}{t}=\xi= \sqrt{\frac{k(k+{\tilde c})}{4\beta(1-\beta)}}\,\qquad\mbox{and}\qquad\vv=\sqrt{\frac{k}{k+{\tilde c}}\frac{1-\beta}{\beta}}\,,
\label{sca-def}
\ee
is an attractor solution of the VOS equations, in the case of a decelerating power-law expansion of the universe --- with $a\propto t^\beta$ and $0<\beta<1$. (For a detailed discussion of the scaling solutions of cosmic string and other p-brane networks, see \cite{Sousa:2011ew,Sousa:2011iu,Avelino:2012qy}.) Note however that such solutions are only possible deep into the matter and radiation epochs. During the radiation-matter transition, the network enters a long-lasting transitional period during which it is not in a linear scaling regime \cite{Avelino:2012qy,Sousa:2013aaa}. Note also that the matter era might not be long enough for the network to reestablish scale-invariant evolution before the onset of dark energy. During a phase of accelerated expansion, the network is conformally stretched \cite{Sousa:2011ew} with $L\propto a$ and $\vv\to0$. Cosmic string networks are, then, diluted away rapidly by the accelerated expansion of the universe once it becomes dark-energy-dominated.

\section{Stochastic gravitational wave background in the Small-loop Regime\label{SGWB}}

The creation of loops is expected to occur copiously throughout the evolution of cosmic string networks. Once a loop detaches from the long string network, it oscillates relativistically and decays in the form of GWs. There are, at any given time, several cosmic string loops emitting GWs in different directions. The superimposition of these emissions generates a SGWB, with a characteristic shape, spanning a wide range of frequencies \cite{Vilenkin:1981bx,Hogan:1984is,Brandenberger:1986xn,Accetta:1988bg}. Cosmic string loops emit GWs in a discrete set of frequencies $f^{\rm em}_j=2j/l$ determined by their physical length $l$ at the time of emission ($f_j^{\rm em}$ is the frequency of the $j$-th harmonic mode). We shall start by assuming that the loops emit all their energy in a single harmonic mode $j$. Although this is not a realistic assumption ---   considering higher order modes has, indeed, significant impact on the spectrum \cite{Olmez:2010bi,Sanidas:2012ee,Sousa:2013aaa} --- we shall recover the full result in the end of this section.

It is often assumed that string loops are created with a size that is a fixed fraction of the characteristic length of the network at the time of birth ($t_b$)
\be
l_b=\alpha L(t_b)\,,
\ee
where $\alpha$ is a constant parameter. Loops lose energy at a constant rate, $dE/dt=\Gamma G\mu^2\,$, and thus their length decreases as GWs are emitted:
\be
l(t)=\alpha L(t_b)-\Gamma G\mu (t-t_b)\,,
\label{loopsize}
\ee
for $t_b<t<t_d$, where $t_d$ is the time of death of the loop, $\Gamma \sim 65$ \cite{Vilenkin:1981bx,Quashnock:1990wv} is a parameter characterizing the efficiency of GW emission, and $G$ is the gravitational constant.

Several studies \cite{Siemens:2001dx,Siemens:2002dj,Polchinski:2006ee,Siemens:2006yp,Polchinski:2007rg,Copeland:2009dk} suggest that cosmic string loops are created with a typical lengthscale that is smaller than the gravitational back-reaction scale, $\Gamma G\mu$ (referred to as small loops). This question, however, is not settled: while some studies indicate that loop size may be closer to the Hubble radius (with $\alpha \sim 10^{-1}-10^{-3}$) \cite{Martins:2005es,Ringeval:2005kr,Vanchurin:2005pa,Olum:2006ix}, others suggest microscopic loops, with a typical length similar to string thickness \cite{Vincent:1996qr,Vincent:1997cx,Bevis:2007gh}. There is also work that favors considerable loop production at significantly different scales \cite{Vanchurin:2007ee,Lorenz:2010sm}. In this paper, we shall focus on the small-loop regime ($\alpha\ll\Gamma G\mu$).

In the small-loop regime, loops live less than a Hubble time, $t_H=H^{-1}$. It is, therefore, reasonable to assume that their energy is radiated in GWs immediately after formation. This energy, however, is not radiated in a single frequency: as the loop size decreases, the GW frequency must increase. In this case, though, this occurs effectively instantaneously in the cosmological timescale. Thus, if the size of the loop at the moment of creation is $l(t)$, it radiates GWs with frequencies
\be
f>f_{\rm min}=\frac{2j}{l(t)}\frac{a(t)}{a_0}\,,
\ee
at the present time. (We use the subscript `0' to refer to the value of the parameter at the present time.)

The amplitude of the SGWB is often quantified by the energy density in GWs, $\rho_{\rm GW}$, per logarithmic frequency interval in units of critical density ($\rho_{\rm c}=(8\pi G) /(3H_0^2)$):
\be
\Omega_{\rm GW}=\frac{1}{\rho_{\rm c}}\frac{d\rho_{\rm GW}}{d\log f}\,.
\ee
The distribution of the power radiated by small loops over the different frequencies is described by the following probability distribution function
\be
p(f)=p(l)\left|\frac{dl}{df}\right| \Theta(f-f_{\rm min})=\frac{f_{\rm min}}{f^2}\Theta (f-f_{\rm min})\,,
\label{pdf}
\ee
where $\Theta(f-f_{\rm min})=1$, for $f\ge f_{\rm min}$, and vanishes for all other $f$. In deriving Eq. (\ref{pdf}), we 
used $dl/df=-2j/f^2$ and the fact that $p(l)=dE/dl$ is constant. Hence
\be
\left. \frac{d\rho_{\rm GW}}{dtdf}\right|_{\rm loops}=\left.\frac{d\rho}{dt}\right|_{\rm loops}\left(\frac{a(t)}{a_0}\right)^{4}\frac{f_{\rm min}}{f^2}\Theta(f-f_{\min})\,,
\ee
where $d\rho/dt |_{\rm loops}$ is given by Eq. (\ref{loss}), and the term dependent on $a(t)$ accounts for the dilution of $\rho_{\rm GW}$ caused by the background expansion. The SGWB spectrum may, then, be computed as follows
\be
\Omega_{\rm gw}^j(f)=\frac{16j\pi G}{3H_0^2 a_0^5}\int_{t_ {\rm min}}^{t_0}\left.\frac{d\rho}{dt}\right|_{\rm loops}\frac{a^5(t)}{\alpha f L(t)}dt \,,
\label{omegasmall}
\ee
where $t_{\rm min}$ is the time of creation of the loops that have $f_{\rm min}=f$.

Gravitational back-reaction damps modes with higher frequencies more efficiently than it does low-frequency modes \cite{Battye:1994qa,Battye:1997ji}. The power emitted in each modes is
\be
\frac{dE_j}{dt}=G\mu^2 \frac{\Gamma}{\mathcal{E}}j^{-q}\,,
\ee
where $\mathcal{E}=\sum_m^{n_s}m^{-q}$ and $q$ a parameter that depends on the shape of the loops. It has been shown that $q \approx 2$ for loops with kinks and $q \approx 4/3$ for cuspy loops \cite{vilenkin2000cosmic}. Here, we have also introduced a cut-off, $n_s$, to the summation in $\mathcal{E}$. Previous work \cite{Sanidas:2012ee,Sousa:2013aaa} has shown that it is sufficient to consider modes up to $n_s=10^3(10^5)$ for loops with kinks (cusps): the spectrum remains unchanged by the inclusion of higher order modes. The full SGWB spectrum may then be obtained by performing a weighed summation of the spectra associated with the different harmonic modes:
\be
\Omega_{\rm GW}(f)=\sum_j^{n_s}\frac{j^{-q}}{\mathcal{E}}\Omega_{\rm GW}^j(f)\,.
\ee

Note that this method to compute the SGWB power spectrum is only valid in the small-loop regime. However, in this regime, it produces identical results to standard methods with the advantage of requiring significantly less computation time. We refer the reader to \cite{Sousa:2013aaa} (see also \cite{Sanidas:2012ee,Kuroyanagi:2012wm}) for a more general analysis. 

\section{Analytic approximation to the stochastic gravitational wave background spectrum\label{approx}}

\begin{figure}
\centering
\includegraphics[width=3.4in]{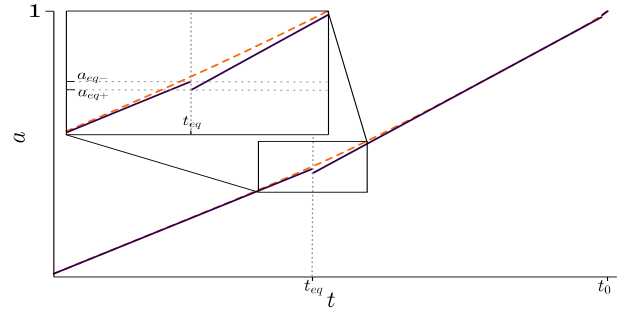}
\caption{Time evolution of the scale factor (dashed line) and of the corresponding fitting function, $a_{\rm fit}(t)$ (solid line). Here, we took $h=0.673$, $\Omega_{\Lambda}^0=0.685$, $\Omega_m^0=0.315$ and $z_{eq}=3391$, consistently with the Planck data combined with the WMAP polarization 9-year data \cite{Ade:2013zuv}. We also took $a_0=1$, and chose a logarithmic scale for both axes.}
\label{afit}
\end{figure}

In this section, we shall compute an analytical approximation to the SGWB spectrum in the small-loop regime. Firstly, we shall assume that the string network exhibits scale-invariant evolution throughout the cosmological history. This assumption --- despite not being very realistic --- is very common in computations of the SGWB spectrum and the effects on its shape have been discussed in \cite{Sousa:2013aaa}. Let $\vv_r$ ($\vv_m$) and $\xi_r$ ($\xi_m$) be the scaling constants that characterize $\vv$ and $L$ during the radiation (matter) era. We shall assume that the transition between these values occurs in a step-like manner at the time of radiation-matter equality, $t_{eq}$. We shall also assume that the universe contains radiation, matter and a cosmological constant ($\Lambda$) and that the evolution of the  scale factor is determined by the dominant component of the energy density. Under this assumption, we have that

\be
a_{\rm fit}(t)=\left\{ \begin{array}{ll}
                    a_{eq-}(t/t_{eq})^{1/2}\,, & \mbox{for $t<t_{eq}$}\\
                    a_{eq+}(t/t_{eq})^{2/3}\,, & \mbox{for $t_{eq}<t<t_{\Lambda}$}\\
                    a_0 \exp[H_0(t-t_0)]\,, & \mbox{for $t_{\Lambda}<t$}\label{afiteq}
                   \end{array}\right. \,,
\ee
where $t_{\Lambda}$ is the instant of time when the energy densities of matter and $\Lambda$ are equal. Note that Eq. (\ref{afiteq}) must be discontinuous at $t_{eq}$ and $t_{\Lambda}$ in order to adjust to the realistic evolution of the scale factor deep in the radiation, matter and $\Lambda$ eras. We have used the value of the scale factor in an instant deep in the radiation era ($a(t_r)=a_r$) and another in the matter era ($a(t_m)=a_m$) to determine the constants $a_{eq-}=a_r(t_{eq}/t_r)^{1/2}$ and $a_{eq+}=a_m(t_{eq}/t_m)^{2/3}$. In Fig. \ref{afit}, the resulting fitting function and the complete evolution of  $a(t)$ are plotted alongside.

We then obtain the following analytical approximation to the SGWB spectrum:

\begin{widetext}
\be
\Omega_{\rm gw}^j h^2(f) = \mathcal{K} \left\{\frac{\vv_r}{\xi_r^3}\frac{\alpha}{j t_{eq}^2}\left(\frac{a_{eq-}}{a_0}\right)^4+
         \frac{3}{f}\frac{\vv_m}{\xi_m^4 t_{\Lambda}^3}\left(\frac{a_{\Lambda-}}{a_0}\right)^5\left[1-\frac{2j}{\alpha\xi_m t_{\Lambda}f}\left(\frac{a_{\Lambda-}}{a_0}\right)\right] \right\}\,,\quad\mbox{for}\quad f\ge\frac{2ja_{\Lambda-}}{t_{\Lambda}} \,,
\label{specrad}
\ee
\end{widetext}
where we have defined the constants $\mathcal{K}=(16j\pi G\mu h^2{\tilde c})/(3H_0^2\alpha)$, and $a_{\Lambda-}=a_{eq+}(t_{\Lambda}/t_{eq})^{2/3}$. Note that our simplifying assumptions --- the abrupt change in $a(t)$, $\vv$ and $L$ at $t_{eq}$ --- give rise to an additional unphysical term which we neglected in the derivation of Eq. (\ref{specrad}). Note also that the contribution of the loops created after $t_{\Lambda}$ was not included in this expression. Nevertheless, once the universe becomes dark-energy-dominated, the network starts being conformally stretched. The amount of energy that is lost due to loop production in this regime decreases steeply (see Eq. (\ref{loss})).
 
The constant high-frequency portion of the SGWB is created by small loops that decay during the radiation era --- whose radiation is, thus, highly redshifted --- and its amplitude is determined by the first term in Eq. (\ref{specrad}). As one moves towards smaller frequencies, the SGWB then starts receiving contributions from loops created during the matter era, which causes the spectral density to increase almost linearly. However, the spectrum develops a peak in the low-frequency range and, as $f$ increases, it slopes approximately linearly towards the radiation era amplitude. This peak is located at $f=4ja_{\Lambda-}/(\alpha \xi_m t_{\Lambda} a_0)$ and its existence is caused by the suppression of GW emission at $t_{\Lambda}$.  Using Eq (\ref{specrad}), one finds that the relative amplitude of the peak of the spectrum when compared to the flat portion is
\be
\frac{\Omega_{\rm peak}}{\Omega_{\rm rad}}=\frac{3}{8}\frac{\bar{v}_r}{\bar{v}_m}\left(\frac{\xi_r}{\xi_m}\right)^3\left(\frac{t_{eq}}{t_{\Lambda}}\right)^{2}\left(\frac{a_{\Lambda-}}{a_{eq-}}\right)^4\,.
\ee

\begin{figure}
\includegraphics[width=3.4in]{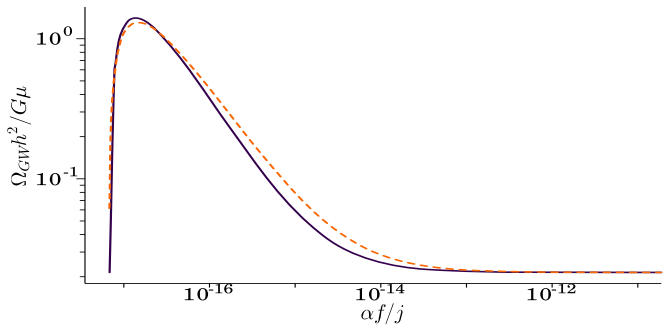}
\caption{The stochastic gravitational wave background generated by a cosmic string network described by the VOS model (dashed line), and the corresponding analytical approximation (solid line). Here, we took ${\tilde c}=0.23$, and the values of the scaling constants ($\xi_m$, $\xi_r$, $\vv_m$ and $\vv_r$) were determined by solving Eqs. (\ref{kpar}) and (\ref{sca-def}). We have also used the cosmological parameters from the Planck mission \cite{Ade:2013zuv}.}
\label{ana}
\end{figure}

The shape of the SGWB spectrum is highly dependent on $\alpha$, and $G\mu$ (see e.g. \cite{Sanidas:2012ee,Kuroyanagi:2012wm,Sousa:2013aaa}). Changing these parameters affects the way the energy density of loops is radiated as a function of time.  Still, if loops radiate their energy rapidly after formation as is the case when $\Gamma G\mu>\alpha$, the rescaled spectrum is essentially the same for all $\alpha$ and $G\mu$. Changing $\alpha$, however, shifts the frequency of the emitted radiation and the amplitude of the spectrum increases linearly with $G\mu$. Fig. \ref{ana} shows the SGWB spectrum generated by a cosmic string network obtained using the VOS model, in units of $G\mu$, as a function of $\alpha f/j$, alongside the analytical approximation given in Eq. (\ref{specrad}). This plot shows that our approximation successfully predicts the amplitude of the flat part of the spectrum and the location and amplitude of the peak of the spectrum. Our approximation slightly underestimates the amplitude of the power spectrum in the mid-frequency range corresponding to the radiation-matter transition and the onset of matter-domination. This is a consequence of the assumption of scale-invariant network evolution: it has been shown in \cite{Sousa:2013aaa} that this assumption causes an underestimation of the number of loops produced during most of the matter era, leading to a slightly narrower peak. Remarkably, despite the simplifying assumptions necessary to make the problem tractable analytically, our  approximation provides an excellent fit to the SGWB spectrum generated by small loops.

\section{Conclusions\label{conc}}

We have proposed an alternative method to compute the SGWB power spectrum generated by cosmic strings, in the small-loop regime. This method does not require underlying simplifications regarding cosmic string network evolution --- avoiding the common assumption of scale-invariance --- thus allowing for an efficient computation of the spectrum generated by string networks undergoing a realistic cosmological evolution. Our method is very useful in the small-loop regime where it is much more efficient than standard methods. This is an important advantage since multiple computations of the spectrum covering a multi-parameter space are often necessary to confront different cosmic string scenarios with the observational data.

Moreover, we used this method to derive an analytical approximation to the SGWB spectrum, which was shown to provide an excellent fit to more elaborate results obtained using the VOS model, in the small-loop regime, over a wide frequency range. This analytical approximation constitutes a useful tool for a first estimation of the SGWB power spectrum generated by cosmic string networks, thus allowing for simple estimates of the associated observational constraints in the small-loop regime.

\begin{acknowledgments}
L.S. is supported by Funda\c{c}\~{a}o para a Ci\^{e}ncia e Tecnologia (FCT, Portugal) and by the European Social Fund (POPH/FSE) through the grant SFRH/BPD/76324/2011. P.A. is supported by FCT through the Investigador FCT contract of reference IF/00863/2012 and POPH/FSE (EC) by FEDER funding through the program "Programa Operacional de Factores de Competitividade - COMPETE. P.A. and L.S. are also partially supported by grant PTDC/FIS/111725/2009 (FCT).
\end{acknowledgments}

\bibliography{SGWB2}

\begin{thebibliography}{66}
\expandafter\ifx\csname natexlab\endcsname\relax\def\natexlab#1{#1}\fi
\expandafter\ifx\csname bibnamefont\endcsname\relax
  \def\bibnamefont#1{#1}\fi
\expandafter\ifx\csname bibfnamefont\endcsname\relax
  \def\bibfnamefont#1{#1}\fi
\expandafter\ifx\csname citenamefont\endcsname\relax
  \def\citenamefont#1{#1}\fi
\expandafter\ifx\csname url\endcsname\relax
  \def\url#1{\texttt{#1}}\fi
\expandafter\ifx\csname urlprefix\endcsname\relax\def\urlprefix{URL }\fi
\providecommand{\bibinfo}[2]{#2}
\providecommand{\eprint}[2][]{\url{#2}}

\bibitem[{\citenamefont{Kibble}(1976)}]{Kibble:1976sj}
\bibinfo{author}{\bibfnamefont{T.~W.~B.} \bibnamefont{Kibble}},
  \bibinfo{journal}{J.Phys.} \textbf{\bibinfo{volume}{A9}},
  \bibinfo{pages}{1387} (\bibinfo{year}{1976}).

\bibitem[{\citenamefont{Vilenkin}(1984)}]{Vilenkin:1984ea}
\bibinfo{author}{\bibfnamefont{A.}~\bibnamefont{Vilenkin}},
  \bibinfo{journal}{Astrophys.J.} \textbf{\bibinfo{volume}{282}},
  \bibinfo{pages}{L51} (\bibinfo{year}{1984}).

\bibitem[{\citenamefont{Avelino
  et~al.}(1998{\natexlab{a}})\citenamefont{Avelino, Shellard, Wu, and
  Allen}}]{Avelino:1997hy}
\bibinfo{author}{\bibfnamefont{P.~P.} \bibnamefont{Avelino}},
  \bibinfo{author}{\bibfnamefont{E.~P.~S.} \bibnamefont{Shellard}},
  \bibinfo{author}{\bibfnamefont{J.~H.~P.} \bibnamefont{Wu}}, \bibnamefont{and}
  \bibinfo{author}{\bibfnamefont{B.}~\bibnamefont{Allen}},
  \bibinfo{journal}{Phys.Rev.Lett.} \textbf{\bibinfo{volume}{81}},
  \bibinfo{pages}{2008} (\bibinfo{year}{1998}{\natexlab{a}}),
  \eprint{astro-ph/9712008}.

\bibitem[{\citenamefont{Avelino
  et~al.}(1998{\natexlab{b}})\citenamefont{Avelino, Shellard, Wu, and
  Allen}}]{Avelino:1998vu}
\bibinfo{author}{\bibfnamefont{P.}~\bibnamefont{Avelino}},
  \bibinfo{author}{\bibfnamefont{E.}~\bibnamefont{Shellard}},
  \bibinfo{author}{\bibfnamefont{J.}~\bibnamefont{Wu}}, \bibnamefont{and}
  \bibinfo{author}{\bibfnamefont{B.}~\bibnamefont{Allen}},
  \bibinfo{journal}{Astrophys.J.} \textbf{\bibinfo{volume}{507}},
  \bibinfo{pages}{L101} (\bibinfo{year}{1998}{\natexlab{b}}),
  \eprint{astro-ph/9803120}.

\bibitem[{\citenamefont{Sarangi and Tye}(2002)}]{Sarangi:2002yt}
\bibinfo{author}{\bibfnamefont{S.}~\bibnamefont{Sarangi}} \bibnamefont{and}
  \bibinfo{author}{\bibfnamefont{S.~H.~H.} \bibnamefont{Tye}},
  \bibinfo{journal}{Phys. Lett.} \textbf{\bibinfo{volume}{B536}},
  \bibinfo{pages}{185} (\bibinfo{year}{2002}), \eprint{hep-th/0204074}.

\bibitem[{\citenamefont{Avelino and Liddle}(2004)}]{Avelino:2003nn}
\bibinfo{author}{\bibfnamefont{P.~P.} \bibnamefont{Avelino}} \bibnamefont{and}
  \bibinfo{author}{\bibfnamefont{A.~R.} \bibnamefont{Liddle}},
  \bibinfo{journal}{Mon.Not.Roy.Astron.Soc.} \textbf{\bibinfo{volume}{348}},
  \bibinfo{pages}{105} (\bibinfo{year}{2004}), \eprint{astro-ph/0305357}.

\bibitem[{\citenamefont{Bevis et~al.}(2010)\citenamefont{Bevis, Hindmarsh,
  Kunz, and Urrestilla}}]{Bevis:2010gj}
\bibinfo{author}{\bibfnamefont{N.}~\bibnamefont{Bevis}},
  \bibinfo{author}{\bibfnamefont{M.}~\bibnamefont{Hindmarsh}},
  \bibinfo{author}{\bibfnamefont{M.}~\bibnamefont{Kunz}}, \bibnamefont{and}
  \bibinfo{author}{\bibfnamefont{J.}~\bibnamefont{Urrestilla}},
  \bibinfo{journal}{Phys.Rev.} \textbf{\bibinfo{volume}{D82}},
  \bibinfo{pages}{065004} (\bibinfo{year}{2010}), \eprint{1005.2663}.

\bibitem[{\citenamefont{Vilenkin}(1981)}]{Vilenkin:1981bx}
\bibinfo{author}{\bibfnamefont{A.}~\bibnamefont{Vilenkin}},
  \bibinfo{journal}{Phys.Lett.} \textbf{\bibinfo{volume}{B107}},
  \bibinfo{pages}{47} (\bibinfo{year}{1981}).

\bibitem[{\citenamefont{Hogan and Rees}(1984)}]{Hogan:1984is}
\bibinfo{author}{\bibfnamefont{C.~J.} \bibnamefont{Hogan}} \bibnamefont{and}
  \bibinfo{author}{\bibfnamefont{M.~J.} \bibnamefont{Rees}},
  \bibinfo{journal}{Nature} \textbf{\bibinfo{volume}{311}},
  \bibinfo{pages}{109} (\bibinfo{year}{1984}).

\bibitem[{\citenamefont{Brandenberger et~al.}(1986)\citenamefont{Brandenberger,
  Albrecht, and Turok}}]{Brandenberger:1986xn}
\bibinfo{author}{\bibfnamefont{R.~H.} \bibnamefont{Brandenberger}},
  \bibinfo{author}{\bibfnamefont{A.}~\bibnamefont{Albrecht}}, \bibnamefont{and}
  \bibinfo{author}{\bibfnamefont{N.}~\bibnamefont{Turok}},
  \bibinfo{journal}{Nucl.Phys.} \textbf{\bibinfo{volume}{B277}},
  \bibinfo{pages}{605} (\bibinfo{year}{1986}).

\bibitem[{\citenamefont{Accetta and Krauss}(1989)}]{Accetta:1988bg}
\bibinfo{author}{\bibfnamefont{F.~S.} \bibnamefont{Accetta}} \bibnamefont{and}
  \bibinfo{author}{\bibfnamefont{L.~M.} \bibnamefont{Krauss}},
  \bibinfo{journal}{Nucl.Phys.} \textbf{\bibinfo{volume}{B319}},
  \bibinfo{pages}{747} (\bibinfo{year}{1989}).

\bibitem[{\citenamefont{Sigg}(2008)}]{Sigg:2008zz}
\bibinfo{author}{\bibfnamefont{D.}~\bibnamefont{Sigg}}
  (\bibinfo{collaboration}{LIGO Scientific Collaboration}),
  \bibinfo{journal}{Class.Quant.Grav.} \textbf{\bibinfo{volume}{25}},
  \bibinfo{pages}{114041} (\bibinfo{year}{2008}).

\bibitem[{\citenamefont{Accadia et~al.}(2011)\citenamefont{Accadia, Acernese,
  Antonucci, Astone, Ballardin et~al.}}]{Accadia:2011zzc}
\bibinfo{author}{\bibfnamefont{T.}~\bibnamefont{Accadia}},
  \bibinfo{author}{\bibfnamefont{F.}~\bibnamefont{Acernese}},
  \bibinfo{author}{\bibfnamefont{F.}~\bibnamefont{Antonucci}},
  \bibinfo{author}{\bibfnamefont{P.}~\bibnamefont{Astone}},
  \bibinfo{author}{\bibfnamefont{G.}~\bibnamefont{Ballardin}},
  \bibnamefont{et~al.}, \bibinfo{journal}{Class.Quant.Grav.}
  \textbf{\bibinfo{volume}{28}}, \bibinfo{pages}{114002}
  (\bibinfo{year}{2011}).

\bibitem[{\citenamefont{Kuroda}(2010)}]{Kuroda:2010zzb}
\bibinfo{author}{\bibfnamefont{K.}~\bibnamefont{Kuroda}}
  (\bibinfo{collaboration}{LCGT Collaboration}),
  \bibinfo{journal}{Class.Quant.Grav.} \textbf{\bibinfo{volume}{27}},
  \bibinfo{pages}{084004} (\bibinfo{year}{2010}).

\bibitem[{\citenamefont{Amaro-Seoane et~al.}(2012)\citenamefont{Amaro-Seoane,
  Aoudia, Babak, Binetruy, Berti et~al.}}]{AmaroSeoane:2012km}
\bibinfo{author}{\bibfnamefont{P.}~\bibnamefont{Amaro-Seoane}},
  \bibinfo{author}{\bibfnamefont{S.}~\bibnamefont{Aoudia}},
  \bibinfo{author}{\bibfnamefont{S.}~\bibnamefont{Babak}},
  \bibinfo{author}{\bibfnamefont{P.}~\bibnamefont{Binetruy}},
  \bibinfo{author}{\bibfnamefont{E.}~\bibnamefont{Berti}}, \bibnamefont{et~al.}
  (\bibinfo{year}{2012}), \eprint{1201.3621}.

\bibitem[{\citenamefont{Kawamura et~al.}(2011)\citenamefont{Kawamura, Ando,
  Seto, Sato, Nakamura et~al.}}]{Kawamura:2011zz}
\bibinfo{author}{\bibfnamefont{S.}~\bibnamefont{Kawamura}},
  \bibinfo{author}{\bibfnamefont{M.}~\bibnamefont{Ando}},
  \bibinfo{author}{\bibfnamefont{N.}~\bibnamefont{Seto}},
  \bibinfo{author}{\bibfnamefont{S.}~\bibnamefont{Sato}},
  \bibinfo{author}{\bibfnamefont{T.}~\bibnamefont{Nakamura}},
  \bibnamefont{et~al.}, \bibinfo{journal}{Class.Quant.Grav.}
  \textbf{\bibinfo{volume}{28}}, \bibinfo{pages}{094011}
  (\bibinfo{year}{2011}).

\bibitem[{\citenamefont{Manchester}(2008)}]{Manchester:2007mx}
\bibinfo{author}{\bibfnamefont{R.~N.} \bibnamefont{Manchester}},
  \bibinfo{journal}{AIP Conf.Proc.} \textbf{\bibinfo{volume}{983}},
  \bibinfo{pages}{584} (\bibinfo{year}{2008}), \eprint{0710.5026}.

\bibitem[{\citenamefont{Ferdman et~al.}(2010)\citenamefont{Ferdman, van
  Haasteren, Bassa, Burgay, Cognard et~al.}}]{Ferdman:2010xq}
\bibinfo{author}{\bibfnamefont{R.~D.} \bibnamefont{Ferdman}},
  \bibinfo{author}{\bibfnamefont{R.}~\bibnamefont{van Haasteren}},
  \bibinfo{author}{\bibfnamefont{C.~G.} \bibnamefont{Bassa}},
  \bibinfo{author}{\bibfnamefont{M.}~\bibnamefont{Burgay}},
  \bibinfo{author}{\bibfnamefont{I.}~\bibnamefont{Cognard}},
  \bibnamefont{et~al.}, \bibinfo{journal}{Class.Quant.Grav.}
  \textbf{\bibinfo{volume}{27}}, \bibinfo{pages}{084014}
  (\bibinfo{year}{2010}), \eprint{1003.3405}.

\bibitem[{\citenamefont{Demorest et~al.}(2013)\citenamefont{Demorest, Ferdman,
  Gonzalez, Nice, Ransom et~al.}}]{Demorest:2012bv}
\bibinfo{author}{\bibfnamefont{P.~B.} \bibnamefont{Demorest}},
  \bibinfo{author}{\bibfnamefont{R.~D.} \bibnamefont{Ferdman}},
  \bibinfo{author}{\bibfnamefont{M.~E.} \bibnamefont{Gonzalez}},
  \bibinfo{author}{\bibfnamefont{D.}~\bibnamefont{Nice}},
  \bibinfo{author}{\bibfnamefont{S.}~\bibnamefont{Ransom}},
  \bibnamefont{et~al.}, \bibinfo{journal}{Astrophys.J.}
  \textbf{\bibinfo{volume}{762}}, \bibinfo{pages}{94} (\bibinfo{year}{2013}),
  \eprint{1201.6641}.

\bibitem[{\citenamefont{Smith et~al.}(2006)\citenamefont{Smith, Pierpaoli, and
  Kamionkowski}}]{Smith:2006nka}
\bibinfo{author}{\bibfnamefont{T.~L.} \bibnamefont{Smith}},
  \bibinfo{author}{\bibfnamefont{E.}~\bibnamefont{Pierpaoli}},
  \bibnamefont{and}
  \bibinfo{author}{\bibfnamefont{M.}~\bibnamefont{Kamionkowski}},
  \bibinfo{journal}{Phys.Rev.Lett.} \textbf{\bibinfo{volume}{97}},
  \bibinfo{pages}{021301} (\bibinfo{year}{2006}), \eprint{astro-ph/0603144}.

\bibitem[{713764()}]{Planck:2006aa}
713764 (\bibinfo{year}{2006}), \eprint{astro-ph/0604069}.

\bibitem[{\citenamefont{Baumann et~al.}(2009)}]{Baumann:2008aq}
\bibinfo{author}{\bibfnamefont{D.}~\bibnamefont{Baumann}} \bibnamefont{et~al.}
  (\bibinfo{collaboration}{CMBPol Study Team}), \bibinfo{journal}{AIP
  Conf.Proc.} \textbf{\bibinfo{volume}{1141}}, \bibinfo{pages}{10}
  (\bibinfo{year}{2009}), \eprint{0811.3919}.

\bibitem[{\citenamefont{Cyburt et~al.}(2005)\citenamefont{Cyburt, Fields,
  Olive, and Skillman}}]{Cyburt:2004yc}
\bibinfo{author}{\bibfnamefont{R.~H.} \bibnamefont{Cyburt}},
  \bibinfo{author}{\bibfnamefont{B.~D.} \bibnamefont{Fields}},
  \bibinfo{author}{\bibfnamefont{K.~A.} \bibnamefont{Olive}}, \bibnamefont{and}
  \bibinfo{author}{\bibfnamefont{E.}~\bibnamefont{Skillman}},
  \bibinfo{journal}{Astropart.Phys.} \textbf{\bibinfo{volume}{23}},
  \bibinfo{pages}{313} (\bibinfo{year}{2005}), \eprint{astro-ph/0408033}.

\bibitem[{\citenamefont{Caldwell and Allen}(1992)}]{Caldwell:1991jj}
\bibinfo{author}{\bibfnamefont{R.~R.} \bibnamefont{Caldwell}} \bibnamefont{and}
  \bibinfo{author}{\bibfnamefont{B.}~\bibnamefont{Allen}},
  \bibinfo{journal}{Phys.Rev.} \textbf{\bibinfo{volume}{D45}},
  \bibinfo{pages}{3447} (\bibinfo{year}{1992}).

\bibitem[{\citenamefont{Caldwell et~al.}(1996)\citenamefont{Caldwell, Battye,
  and Shellard}}]{Caldwell:1996en}
\bibinfo{author}{\bibfnamefont{R.~R.} \bibnamefont{Caldwell}},
  \bibinfo{author}{\bibfnamefont{R.~A.} \bibnamefont{Battye}},
  \bibnamefont{and} \bibinfo{author}{\bibfnamefont{E.~P.~S.}
  \bibnamefont{Shellard}}, \bibinfo{journal}{Phys.Rev.}
  \textbf{\bibinfo{volume}{D54}}, \bibinfo{pages}{7146} (\bibinfo{year}{1996}),
  \eprint{astro-ph/9607130}.

\bibitem[{\citenamefont{Battye et~al.}(1997)\citenamefont{Battye, Caldwell, and
  Shellard}}]{Battye:1997ji}
\bibinfo{author}{\bibfnamefont{R.~A.} \bibnamefont{Battye}},
  \bibinfo{author}{\bibfnamefont{R.~R.} \bibnamefont{Caldwell}},
  \bibnamefont{and} \bibinfo{author}{\bibfnamefont{E.~P.~S.}
  \bibnamefont{Shellard}}, pp. \bibinfo{pages}{11--31} (\bibinfo{year}{1997}),
  \eprint{astro-ph/9706013}.

\bibitem[{\citenamefont{Damour and Vilenkin}(2001)}]{Damour:2001bk}
\bibinfo{author}{\bibfnamefont{T.}~\bibnamefont{Damour}} \bibnamefont{and}
  \bibinfo{author}{\bibfnamefont{A.}~\bibnamefont{Vilenkin}},
  \bibinfo{journal}{Phys.Rev.} \textbf{\bibinfo{volume}{D64}},
  \bibinfo{pages}{064008} (\bibinfo{year}{2001}), \eprint{gr-qc/0104026}.

\bibitem[{\citenamefont{Damour and Vilenkin}(2005)}]{Damour:2004kw}
\bibinfo{author}{\bibfnamefont{T.}~\bibnamefont{Damour}} \bibnamefont{and}
  \bibinfo{author}{\bibfnamefont{A.}~\bibnamefont{Vilenkin}},
  \bibinfo{journal}{Phys.Rev.} \textbf{\bibinfo{volume}{D71}},
  \bibinfo{pages}{063510} (\bibinfo{year}{2005}), \eprint{hep-th/0410222}.

\bibitem[{\citenamefont{Siemens et~al.}(2006)\citenamefont{Siemens, Creighton,
  Maor, Ray~Majumder, Cannon et~al.}}]{Siemens:2006vk}
\bibinfo{author}{\bibfnamefont{X.}~\bibnamefont{Siemens}},
  \bibinfo{author}{\bibfnamefont{J.}~\bibnamefont{Creighton}},
  \bibinfo{author}{\bibfnamefont{I.}~\bibnamefont{Maor}},
  \bibinfo{author}{\bibfnamefont{S.}~\bibnamefont{Ray~Majumder}},
  \bibinfo{author}{\bibfnamefont{K.}~\bibnamefont{Cannon}},
  \bibnamefont{et~al.}, \bibinfo{journal}{Phys.Rev.}
  \textbf{\bibinfo{volume}{D73}}, \bibinfo{pages}{105001}
  (\bibinfo{year}{2006}), \eprint{gr-qc/0603115}.

\bibitem[{\citenamefont{Hogan}(2006)}]{Hogan:2006we}
\bibinfo{author}{\bibfnamefont{C.~J.} \bibnamefont{Hogan}},
  \bibinfo{journal}{Phys.Rev.} \textbf{\bibinfo{volume}{D74}},
  \bibinfo{pages}{043526} (\bibinfo{year}{2006}), \eprint{astro-ph/0605567}.

\bibitem[{\citenamefont{Polchinski and Rocha}(2006)}]{Polchinski:2006ee}
\bibinfo{author}{\bibfnamefont{J.}~\bibnamefont{Polchinski}} \bibnamefont{and}
  \bibinfo{author}{\bibfnamefont{J.~V.} \bibnamefont{Rocha}},
  \bibinfo{journal}{Phys.Rev.} \textbf{\bibinfo{volume}{D74}},
  \bibinfo{pages}{083504} (\bibinfo{year}{2006}), \eprint{hep-ph/0606205}.

\bibitem[{\citenamefont{DePies and Hogan}(2007)}]{DePies:2007bm}
\bibinfo{author}{\bibfnamefont{M.~R.} \bibnamefont{DePies}} \bibnamefont{and}
  \bibinfo{author}{\bibfnamefont{C.~J.} \bibnamefont{Hogan}},
  \bibinfo{journal}{Phys.Rev.} \textbf{\bibinfo{volume}{D75}},
  \bibinfo{pages}{125006} (\bibinfo{year}{2007}), \eprint{astro-ph/0702335}.

\bibitem[{\citenamefont{Olmez et~al.}(2010)\citenamefont{Olmez, Mandic, and
  Siemens}}]{Olmez:2010bi}
\bibinfo{author}{\bibfnamefont{S.}~\bibnamefont{Olmez}},
  \bibinfo{author}{\bibfnamefont{V.}~\bibnamefont{Mandic}}, \bibnamefont{and}
  \bibinfo{author}{\bibfnamefont{X.}~\bibnamefont{Siemens}},
  \bibinfo{journal}{Phys.Rev.} \textbf{\bibinfo{volume}{D81}},
  \bibinfo{pages}{104028} (\bibinfo{year}{2010}), \eprint{1004.0890}.

\bibitem[{\citenamefont{Binetruy et~al.}(2012)\citenamefont{Binetruy, Bohe,
  Caprini, and Dufaux}}]{Binetruy:2012ze}
\bibinfo{author}{\bibfnamefont{P.}~\bibnamefont{Binetruy}},
  \bibinfo{author}{\bibfnamefont{A.}~\bibnamefont{Bohe}},
  \bibinfo{author}{\bibfnamefont{C.}~\bibnamefont{Caprini}}, \bibnamefont{and}
  \bibinfo{author}{\bibfnamefont{J.-F.} \bibnamefont{Dufaux}},
  \bibinfo{journal}{JCAP} \textbf{\bibinfo{volume}{1206}}, \bibinfo{pages}{027}
  (\bibinfo{year}{2012}), \eprint{1201.0983}.

\bibitem[{\citenamefont{Kuroyanagi et~al.}(2012)\citenamefont{Kuroyanagi,
  Miyamoto, Sekiguchi, Takahashi, and Silk}}]{Kuroyanagi:2012wm}
\bibinfo{author}{\bibfnamefont{S.}~\bibnamefont{Kuroyanagi}},
  \bibinfo{author}{\bibfnamefont{K.}~\bibnamefont{Miyamoto}},
  \bibinfo{author}{\bibfnamefont{T.}~\bibnamefont{Sekiguchi}},
  \bibinfo{author}{\bibfnamefont{K.}~\bibnamefont{Takahashi}},
  \bibnamefont{and} \bibinfo{author}{\bibfnamefont{J.}~\bibnamefont{Silk}},
  \bibinfo{journal}{Phys.Rev.} \textbf{\bibinfo{volume}{D86}},
  \bibinfo{pages}{023503} (\bibinfo{year}{2012}), \eprint{1202.3032}.

\bibitem[{\citenamefont{Sanidas et~al.}(2012)\citenamefont{Sanidas, Battye, and
  Stappers}}]{Sanidas:2012ee}
\bibinfo{author}{\bibfnamefont{S.~A.} \bibnamefont{Sanidas}},
  \bibinfo{author}{\bibfnamefont{R.~A.} \bibnamefont{Battye}},
  \bibnamefont{and} \bibinfo{author}{\bibfnamefont{B.~W.}
  \bibnamefont{Stappers}}, \bibinfo{journal}{Phys.Rev.}
  \textbf{\bibinfo{volume}{D85}}, \bibinfo{pages}{122003}
  (\bibinfo{year}{2012}), \eprint{1201.2419}.

\bibitem[{\citenamefont{Sousa and Avelino}(2013)}]{Sousa:2013aaa}
\bibinfo{author}{\bibfnamefont{L.}~\bibnamefont{Sousa}} \bibnamefont{and}
  \bibinfo{author}{\bibfnamefont{P.~P.} \bibnamefont{Avelino}},
  \bibinfo{journal}{Phys.Rev.} \textbf{\bibinfo{volume}{D88}},
  \bibinfo{pages}{023516} (\bibinfo{year}{2013}), \eprint{1304.2445}.

\bibitem[{\citenamefont{Martins and Shellard}(1996)}]{Martins:1996jp}
\bibinfo{author}{\bibfnamefont{C.~J. A.~P.} \bibnamefont{Martins}}
  \bibnamefont{and} \bibinfo{author}{\bibfnamefont{E.~P.~S.}
  \bibnamefont{Shellard}}, \bibinfo{journal}{Phys.Rev.}
  \textbf{\bibinfo{volume}{D54}}, \bibinfo{pages}{2535} (\bibinfo{year}{1996}),
  \eprint{hep-ph/9602271}.

\bibitem[{\citenamefont{Martins and Shellard}(2002)}]{Martins:2000cs}
\bibinfo{author}{\bibfnamefont{C.~J. A.~P.} \bibnamefont{Martins}}
  \bibnamefont{and} \bibinfo{author}{\bibfnamefont{E.~P.~S.}
  \bibnamefont{Shellard}}, \bibinfo{journal}{Phys. Rev.}
  \textbf{\bibinfo{volume}{D65}}, \bibinfo{pages}{043514}
  (\bibinfo{year}{2002}), \eprint{hep-ph/0003298}.

\bibitem[{\citenamefont{Sousa and Avelino}(2011{\natexlab{a}})}]{Sousa:2011ew}
\bibinfo{author}{\bibfnamefont{L.}~\bibnamefont{Sousa}} \bibnamefont{and}
  \bibinfo{author}{\bibfnamefont{P.~P.} \bibnamefont{Avelino}},
  \bibinfo{journal}{Phys.Rev.} \textbf{\bibinfo{volume}{D83}},
  \bibinfo{pages}{103507} (\bibinfo{year}{2011}{\natexlab{a}}),
  \eprint{1103.1381}.

\bibitem[{\citenamefont{Sousa and Avelino}(2011{\natexlab{b}})}]{Sousa:2011iu}
\bibinfo{author}{\bibfnamefont{L.}~\bibnamefont{Sousa}} \bibnamefont{and}
  \bibinfo{author}{\bibfnamefont{P.~P.} \bibnamefont{Avelino}},
  \bibinfo{journal}{Phys.Rev.} \textbf{\bibinfo{volume}{D84}},
  \bibinfo{pages}{063502} (\bibinfo{year}{2011}{\natexlab{b}}),
  \eprint{1107.4582}.

\bibitem[{\citenamefont{Kibble}(1985)}]{Kibble:1984hp}
\bibinfo{author}{\bibfnamefont{T.~W.~B.} \bibnamefont{Kibble}},
  \bibinfo{journal}{Nucl. Phys.} \textbf{\bibinfo{volume}{B252}},
  \bibinfo{pages}{227} (\bibinfo{year}{1985}).

\bibitem[{\citenamefont{Bennett and Bouchet}(1988)}]{Bennett:1987vf}
\bibinfo{author}{\bibfnamefont{D.~P.} \bibnamefont{Bennett}} \bibnamefont{and}
  \bibinfo{author}{\bibfnamefont{F.~R.} \bibnamefont{Bouchet}},
  \bibinfo{journal}{Phys.Rev.Lett.} \textbf{\bibinfo{volume}{60}},
  \bibinfo{pages}{257} (\bibinfo{year}{1988}).

\bibitem[{\citenamefont{Albrecht and Turok}(1989)}]{Albrecht:1989mk}
\bibinfo{author}{\bibfnamefont{A.}~\bibnamefont{Albrecht}} \bibnamefont{and}
  \bibinfo{author}{\bibfnamefont{N.}~\bibnamefont{Turok}},
  \bibinfo{journal}{Phys.Rev.} \textbf{\bibinfo{volume}{D40}},
  \bibinfo{pages}{973} (\bibinfo{year}{1989}).

\bibitem[{\citenamefont{Allen and Shellard}(1990)}]{Allen:1990tv}
\bibinfo{author}{\bibfnamefont{B.}~\bibnamefont{Allen}} \bibnamefont{and}
  \bibinfo{author}{\bibfnamefont{E.~P.~S.} \bibnamefont{Shellard}},
  \bibinfo{journal}{Phys.Rev.Lett.} \textbf{\bibinfo{volume}{64}},
  \bibinfo{pages}{119} (\bibinfo{year}{1990}).

\bibitem[{\citenamefont{Copeland et~al.}(1992)\citenamefont{Copeland, Kibble,
  and Austin}}]{Copeland:1991kz}
\bibinfo{author}{\bibfnamefont{E.~J.} \bibnamefont{Copeland}},
  \bibinfo{author}{\bibfnamefont{T.~W.~B.} \bibnamefont{Kibble}},
  \bibnamefont{and} \bibinfo{author}{\bibfnamefont{D.}~\bibnamefont{Austin}},
  \bibinfo{journal}{Phys.Rev.} \textbf{\bibinfo{volume}{D45}},
  \bibinfo{pages}{1000} (\bibinfo{year}{1992}).

\bibitem[{\citenamefont{Vincent
  et~al.}(1997{\natexlab{a}})\citenamefont{Vincent, Hindmarsh, and
  Sakellariadou}}]{Vincent:1996rb}
\bibinfo{author}{\bibfnamefont{G.~R.} \bibnamefont{Vincent}},
  \bibinfo{author}{\bibfnamefont{M.}~\bibnamefont{Hindmarsh}},
  \bibnamefont{and}
  \bibinfo{author}{\bibfnamefont{M.}~\bibnamefont{Sakellariadou}},
  \bibinfo{journal}{Phys.Rev.} \textbf{\bibinfo{volume}{D56}},
  \bibinfo{pages}{637} (\bibinfo{year}{1997}{\natexlab{a}}),
  \eprint{astro-ph/9612135}.

\bibitem[{\citenamefont{Avelino and Sousa}(2012)}]{Avelino:2012qy}
\bibinfo{author}{\bibfnamefont{P.~P.} \bibnamefont{Avelino}} \bibnamefont{and}
  \bibinfo{author}{\bibfnamefont{L.}~\bibnamefont{Sousa}},
  \bibinfo{journal}{Phys.Rev.} \textbf{\bibinfo{volume}{D85}},
  \bibinfo{pages}{083525} (\bibinfo{year}{2012}), \eprint{1202.6298}.

\bibitem[{\citenamefont{Quashnock and Spergel}(1990)}]{Quashnock:1990wv}
\bibinfo{author}{\bibfnamefont{J.~M.} \bibnamefont{Quashnock}}
  \bibnamefont{and} \bibinfo{author}{\bibfnamefont{D.~N.}
  \bibnamefont{Spergel}}, \bibinfo{journal}{Phys.Rev.}
  \textbf{\bibinfo{volume}{D42}}, \bibinfo{pages}{2505} (\bibinfo{year}{1990}).

\bibitem[{\citenamefont{Siemens and Olum}(2001)}]{Siemens:2001dx}
\bibinfo{author}{\bibfnamefont{X.}~\bibnamefont{Siemens}} \bibnamefont{and}
  \bibinfo{author}{\bibfnamefont{K.~D.} \bibnamefont{Olum}},
  \bibinfo{journal}{Nucl.Phys.} \textbf{\bibinfo{volume}{B611}},
  \bibinfo{pages}{125} (\bibinfo{year}{2001}), \eprint{gr-qc/0104085}.

\bibitem[{\citenamefont{Siemens et~al.}(2002)\citenamefont{Siemens, Olum, and
  Vilenkin}}]{Siemens:2002dj}
\bibinfo{author}{\bibfnamefont{X.}~\bibnamefont{Siemens}},
  \bibinfo{author}{\bibfnamefont{K.~D.} \bibnamefont{Olum}}, \bibnamefont{and}
  \bibinfo{author}{\bibfnamefont{A.}~\bibnamefont{Vilenkin}},
  \bibinfo{journal}{Phys.Rev.} \textbf{\bibinfo{volume}{D66}},
  \bibinfo{pages}{043501} (\bibinfo{year}{2002}), \eprint{gr-qc/0203006}.

\bibitem[{\citenamefont{Siemens et~al.}(2007)\citenamefont{Siemens, Mandic, and
  Creighton}}]{Siemens:2006yp}
\bibinfo{author}{\bibfnamefont{X.}~\bibnamefont{Siemens}},
  \bibinfo{author}{\bibfnamefont{V.}~\bibnamefont{Mandic}}, \bibnamefont{and}
  \bibinfo{author}{\bibfnamefont{J.}~\bibnamefont{Creighton}},
  \bibinfo{journal}{Phys.Rev.Lett.} \textbf{\bibinfo{volume}{98}},
  \bibinfo{pages}{111101} (\bibinfo{year}{2007}), \eprint{astro-ph/0610920}.

\bibitem[{\citenamefont{Polchinski and Rocha}(2007)}]{Polchinski:2007rg}
\bibinfo{author}{\bibfnamefont{J.}~\bibnamefont{Polchinski}} \bibnamefont{and}
  \bibinfo{author}{\bibfnamefont{J.~V.} \bibnamefont{Rocha}},
  \bibinfo{journal}{Phys.Rev.} \textbf{\bibinfo{volume}{D75}},
  \bibinfo{pages}{123503} (\bibinfo{year}{2007}), \eprint{gr-qc/0702055}.

\bibitem[{\citenamefont{Copeland and Kibble}(2009)}]{Copeland:2009dk}
\bibinfo{author}{\bibfnamefont{E.~J.} \bibnamefont{Copeland}} \bibnamefont{and}
  \bibinfo{author}{\bibfnamefont{T.~W.~B.} \bibnamefont{Kibble}},
  \bibinfo{journal}{Phys.Rev.} \textbf{\bibinfo{volume}{D80}},
  \bibinfo{pages}{123523} (\bibinfo{year}{2009}), \eprint{0909.1960}.

\bibitem[{\citenamefont{Martins and Shellard}(2006)}]{Martins:2005es}
\bibinfo{author}{\bibfnamefont{C.~J. A.~P.} \bibnamefont{Martins}}
  \bibnamefont{and} \bibinfo{author}{\bibfnamefont{E.~P.~S.}
  \bibnamefont{Shellard}}, \bibinfo{journal}{Phys.Rev.}
  \textbf{\bibinfo{volume}{D73}}, \bibinfo{pages}{043515}
  (\bibinfo{year}{2006}), \eprint{astro-ph/0511792}.

\bibitem[{\citenamefont{Ringeval et~al.}(2007)\citenamefont{Ringeval,
  Sakellariadou, and Bouchet}}]{Ringeval:2005kr}
\bibinfo{author}{\bibfnamefont{C.}~\bibnamefont{Ringeval}},
  \bibinfo{author}{\bibfnamefont{M.}~\bibnamefont{Sakellariadou}},
  \bibnamefont{and} \bibinfo{author}{\bibfnamefont{F.}~\bibnamefont{Bouchet}},
  \bibinfo{journal}{JCAP} \textbf{\bibinfo{volume}{0702}}, \bibinfo{pages}{023}
  (\bibinfo{year}{2007}), \eprint{astro-ph/0511646}.

\bibitem[{\citenamefont{Vanchurin et~al.}(2006)\citenamefont{Vanchurin, Olum,
  and Vilenkin}}]{Vanchurin:2005pa}
\bibinfo{author}{\bibfnamefont{V.}~\bibnamefont{Vanchurin}},
  \bibinfo{author}{\bibfnamefont{K.~D.} \bibnamefont{Olum}}, \bibnamefont{and}
  \bibinfo{author}{\bibfnamefont{A.}~\bibnamefont{Vilenkin}},
  \bibinfo{journal}{Phys.Rev.} \textbf{\bibinfo{volume}{D74}},
  \bibinfo{pages}{063527} (\bibinfo{year}{2006}), \eprint{gr-qc/0511159}.

\bibitem[{\citenamefont{Olum and Vanchurin}(2007)}]{Olum:2006ix}
\bibinfo{author}{\bibfnamefont{K.~D.} \bibnamefont{Olum}} \bibnamefont{and}
  \bibinfo{author}{\bibfnamefont{V.}~\bibnamefont{Vanchurin}},
  \bibinfo{journal}{Phys.Rev.} \textbf{\bibinfo{volume}{D75}},
  \bibinfo{pages}{063521} (\bibinfo{year}{2007}), \eprint{astro-ph/0610419}.

\bibitem[{\citenamefont{Vincent
  et~al.}(1997{\natexlab{b}})\citenamefont{Vincent, Hindmarsh, and
  Sakellariadou}}]{Vincent:1996qr}
\bibinfo{author}{\bibfnamefont{G.~R.} \bibnamefont{Vincent}},
  \bibinfo{author}{\bibfnamefont{M.}~\bibnamefont{Hindmarsh}},
  \bibnamefont{and}
  \bibinfo{author}{\bibfnamefont{M.}~\bibnamefont{Sakellariadou}},
  \bibinfo{journal}{Phys.Rev.} \textbf{\bibinfo{volume}{D55}},
  \bibinfo{pages}{573} (\bibinfo{year}{1997}{\natexlab{b}}),
  \eprint{astro-ph/9606137}.

\bibitem[{\citenamefont{Vincent et~al.}(1998)\citenamefont{Vincent, Antunes,
  and Hindmarsh}}]{Vincent:1997cx}
\bibinfo{author}{\bibfnamefont{G.}~\bibnamefont{Vincent}},
  \bibinfo{author}{\bibfnamefont{N.~D.} \bibnamefont{Antunes}},
  \bibnamefont{and}
  \bibinfo{author}{\bibfnamefont{M.}~\bibnamefont{Hindmarsh}},
  \bibinfo{journal}{Phys.Rev.Lett.} \textbf{\bibinfo{volume}{80}},
  \bibinfo{pages}{2277} (\bibinfo{year}{1998}), \eprint{hep-ph/9708427}.

\bibitem[{\citenamefont{Bevis et~al.}(2008)\citenamefont{Bevis, Hindmarsh,
  Kunz, and Urrestilla}}]{Bevis:2007gh}
\bibinfo{author}{\bibfnamefont{N.}~\bibnamefont{Bevis}},
  \bibinfo{author}{\bibfnamefont{M.}~\bibnamefont{Hindmarsh}},
  \bibinfo{author}{\bibfnamefont{M.}~\bibnamefont{Kunz}}, \bibnamefont{and}
  \bibinfo{author}{\bibfnamefont{J.}~\bibnamefont{Urrestilla}},
  \bibinfo{journal}{Phys.Rev.Lett.} \textbf{\bibinfo{volume}{100}},
  \bibinfo{pages}{021301} (\bibinfo{year}{2008}), \eprint{astro-ph/0702223}.

\bibitem[{\citenamefont{Vanchurin}(2008)}]{Vanchurin:2007ee}
\bibinfo{author}{\bibfnamefont{V.}~\bibnamefont{Vanchurin}},
  \bibinfo{journal}{Phys.Rev.} \textbf{\bibinfo{volume}{D77}},
  \bibinfo{pages}{063532} (\bibinfo{year}{2008}), \eprint{0712.2236}.

\bibitem[{\citenamefont{Lorenz et~al.}(2010)\citenamefont{Lorenz, Ringeval, and
  Sakellariadou}}]{Lorenz:2010sm}
\bibinfo{author}{\bibfnamefont{L.}~\bibnamefont{Lorenz}},
  \bibinfo{author}{\bibfnamefont{C.}~\bibnamefont{Ringeval}}, \bibnamefont{and}
  \bibinfo{author}{\bibfnamefont{M.}~\bibnamefont{Sakellariadou}},
  \bibinfo{journal}{JCAP} \textbf{\bibinfo{volume}{1010}}, \bibinfo{pages}{003}
  (\bibinfo{year}{2010}), \eprint{1006.0931}.

\bibitem[{\citenamefont{Battye and Shellard}(1995)}]{Battye:1994qa}
\bibinfo{author}{\bibfnamefont{R.~A.} \bibnamefont{Battye}} \bibnamefont{and}
  \bibinfo{author}{\bibfnamefont{E.~P.~S.} \bibnamefont{Shellard}},
  \bibinfo{journal}{Phys.Rev.Lett.} \textbf{\bibinfo{volume}{75}},
  \bibinfo{pages}{4354} (\bibinfo{year}{1995}), \eprint{astro-ph/9408078}.

\bibitem[{\citenamefont{Vilenkin and Shellard}(2000)}]{vilenkin2000cosmic}
\bibinfo{author}{\bibfnamefont{A.}~\bibnamefont{Vilenkin}} \bibnamefont{and}
  \bibinfo{author}{\bibfnamefont{E.~P.~S.} \bibnamefont{Shellard}},
  \emph{\bibinfo{title}{{Cosmic strings and other topological defects}}},
  Cambridge monographs on mathematical physics (\bibinfo{publisher}{Cambridge
  University Press}, \bibinfo{year}{2000}).

\bibitem[{\citenamefont{Ade et~al.}(2013)}]{Ade:2013zuv}
\bibinfo{author}{\bibfnamefont{P.~A.~R.} \bibnamefont{Ade}}
  \bibnamefont{et~al.} (\bibinfo{collaboration}{Planck Collaboration})
  (\bibinfo{year}{2013}), \eprint{1303.5076}.

\end{thebibliography}

\end{document}